\documentclass[number,3p,twocolumn]{elsarticle}

\usepackage[pdftex]{graphicx}

\usepackage{amsmath}
\usepackage{amssymb}

\journal{Chemical Physics}

\begin{document}

\begin{frontmatter}



\title{Directed transport in a ratchet with internal and
chemical freedoms}


\author[label1,label2]{T.~Dittrich}
\author[label1,label2]{N.~A.~Naranjo}

\address[label1]{Depto.\ de F\'\i sica, Universidad Nacional de
Colmbia, Bogot\'a D.C., Colombia}
\address[label2]{CeiBA--Complejidad, Bogot\'a D.C., Colombia}

\begin{abstract}
We consider mechanisms of directed transport in a ratchet model comprising,
besides the external freedom where transport occurs, a chemical freedom that
replaces the familiar external driving by an autonomous dynamics providing
energy input, and an internal freedom representing a functional mode of a
motor molecule. The dependence of the current on various parameters is studied
in numerical simulations of our model. In particular, we point out the r\^ole
of the internal freedom as a buffer between energy input and output of
mechanical work that allows a temporary storage of injected energy and can
contribute to the efficiency of current generation.
\end{abstract}



\end{frontmatter}

\section{Introduction}
\label{intro}
Research on directed transport at the microscopic level has benefitted
enormously from the interaction and mutual stimulation of applied
studies close to experimentally accessible phenomena and theoretical
analyses, the paradigm case being the concept of Brownian motors
\cite{Ast94,BHK94,BRH96,Ast97} as it emerged from the discovery of
motor molecules generating directed transport in the living cell
\cite{Lei94,Spu94}. Notwithstanding, the distance between the two
endeavours remains large. An example is the degree of structural
complexity, reflected in the number of degrees of freedom involved, of
the models considered on either side: While realistic
molecular-dynamics simulations of biomolecules typically include the
full set of hundreds or thousands of nuclear freedoms, plus possibly
some of the surrounding solvent (for a review see, e.g.,
\cite{WD&01}), the key notion of ratchets has been developed
considering mere point particles in one-dimensional periodic
potentials \cite{AB96,Rei02}.

Efforts are under way on both sides to bring these ends closer to one
another. On the one hand, in molecular biophysics, the concept of
functional degree of freedom \cite{TS03,NS06} has been conceived for
large-scale conformational changes on long timescales that are
immediately responsible, e.g., for the catalytic activity of a
protein, and can be identified on basis of objective criteria such as
normal-mode analysis \cite{MG&76,Har84,BK85} to establish a hierarchy
among the modes from slow to fast and from collective (global) to
microscopic. On the other hand, elementary ratchet models are being
endowed one by one with additional freedoms to study their possible
r\^ole in directed transport \cite{Mat02}.

The present work is intended to insert another stepstone from the side
of abstract ratchet models towards biophysical realism, by
incorporating a few crucial features of molecular motors: We devise a
model that includes an internal freedom to represent some
functionally relevant conformational change. The concept of molecular
\textit{combustion} motor is incorporated replacing the familiar
periodic external driving by a chemical freedom as coherent energy
source which maintains the system far from thermal equilibrium
\cite{KB00}. The ubiquitous r\^ole of thermal fluctuations is modeled,
as usually, by random forces. At the same time, we avoid the detailed
reconstruction of any specific system such as ${\rm F}_1$-ATPase
\cite{Kin98,Min03,AM08} or even their reduction to ``mechanical toy
models'' \cite{PO95} which typically still are tailored to represent a
certain (species of) motor molecule and therefore involve some
contingency. This allows us to study the effect of internal freedoms
on general grounds, applicable to a broad class of molecules and even
beyond the biophysical realm. In this way, we take up themes set by
pioneering works such as Kaneko \textit{et al.} \cite{NK03} on
molecular machines with internal freedoms and of Mateos on ratchets
based on non-point particles \cite{Mat02}. At the same time, focussing
on the biological context, we leave aside quantum effects which
probably are of minor relevance for transport in the living cell, yet
may be crucial in artificial nanodevices \cite{KLH05}.

The dynamics of our model is simulated in the underdamped regime,
allowing for complex deterministic motion, and in the presence of
thermal noise, analyzed as to transport mechanisms involving the
internal freedom as essential element. We find evidence, above all,
that it serves as a temporary energy storage which partially decouples
directed transport from the energy source and thus decorrelates the
discrete steps in the external freedom from the likewise approximately
quantized discharges of chemical energy into the system---certainly
desirable features from the point of view of robustness, efficiency,
and versatility of a molecular motor. Preliminary results of this work
have been published (in Spanish) in Ref.~\cite{DN09}.

Our model is motivated and constructed in Sec.~\ref{model}. Section
\ref{dynamics} provides a survey of its dynamical behaviour in
different parameter regimes. Transport properties and the r\^ole of
the internal freedom are discussed in Sec.~\ref{transport}. We
conclude in Sec.~\ref{end} with some remarks on open ends.

\section{Model}
\label{model}
We pretend to construct a minimal model of a molecular combustion
motor that goes beyond the familiar ratchet scheme in two essential
points, (i) it includes an internal freedom to represent a functional
(slow, conformational) mode of the molecule, and (ii) it incorporates
the injection of chemical energy, e.g., through hydrolysis of ATP, as
autonomous dynamics of a chemical degree of freedom corresponding to
the reaction coordinate underlying that process \cite{KB00}. The
formulation of our model is inspired in various respects by the
example of F1-ATPase, a prototypical rotational molecular motor
\cite{Kin98,Min03,AM08}, but does not intend its reconstruction in any
biophysical detail, i.e., is to be considered an impressionist view,
at most, of that molecule.

More precisely, we require the model to comply with the following
conditions:
\renewcommand{\theenumi}{\roman{enumi}}
\begin{enumerate}
\item \textit{External freedom} The coordinate $x_{\rm ex}$ is
equivalent to the single freedom of the usual point-particle ratchets.
It will be subjected to a standard ratchet potential, periodic with 
period $X_{\rm ex} = 2\pi$ but breaking invariance under parity 
$x_{\rm ex} \to -x_{\rm ex}$, which can be considered either as cyclic 
(angle) or as extended. To model the conversion of internal kinetic 
energy into directed transport, it will be coupled to the internal 
freedom, but also to the chemical freedom from which it receives 
coherent energy.
\item \textit{Internal freedom} Following Kaneko \textit{et al.}
\cite{NK03}, we model the internal freedom $x_{\rm in}$ like a
pendulum, with a smooth periodic potential, period $X_{\rm in} = 
2\pi$, which however lacks the intentional asymmetry of the ratchet 
potential proper.
\item \textit{Chemical freedom} In order to reproduce the entropical
bias of the ``combustion'' reaction, we impose a non-zero mean
gradient on the otherwise periodic dependence of the potential on the
chemical coordinate $x_{\rm ch}$, i.e., $V(x_{\rm ex},x_{\rm
in},x_{\rm ch}+X_{\rm ch}) = V(x_{\rm ex},x_{\rm in},x_{\rm ch})
+ E_{\rm ch}$, $E_{\rm ch}$ denoting the net energy gain of the
(hydrolysis etc.) reaction per molecule. To keep the behaviour of this
coordinate close to an approximately periodic energy injection, even
in presence of the backaction from the external freedom, we further
choose a large inertia $m_{\rm ch}$.
\end{enumerate}

\begin{figure}[h!]
\begin{center}
\includegraphics[scale=0.3,angle=0  ]{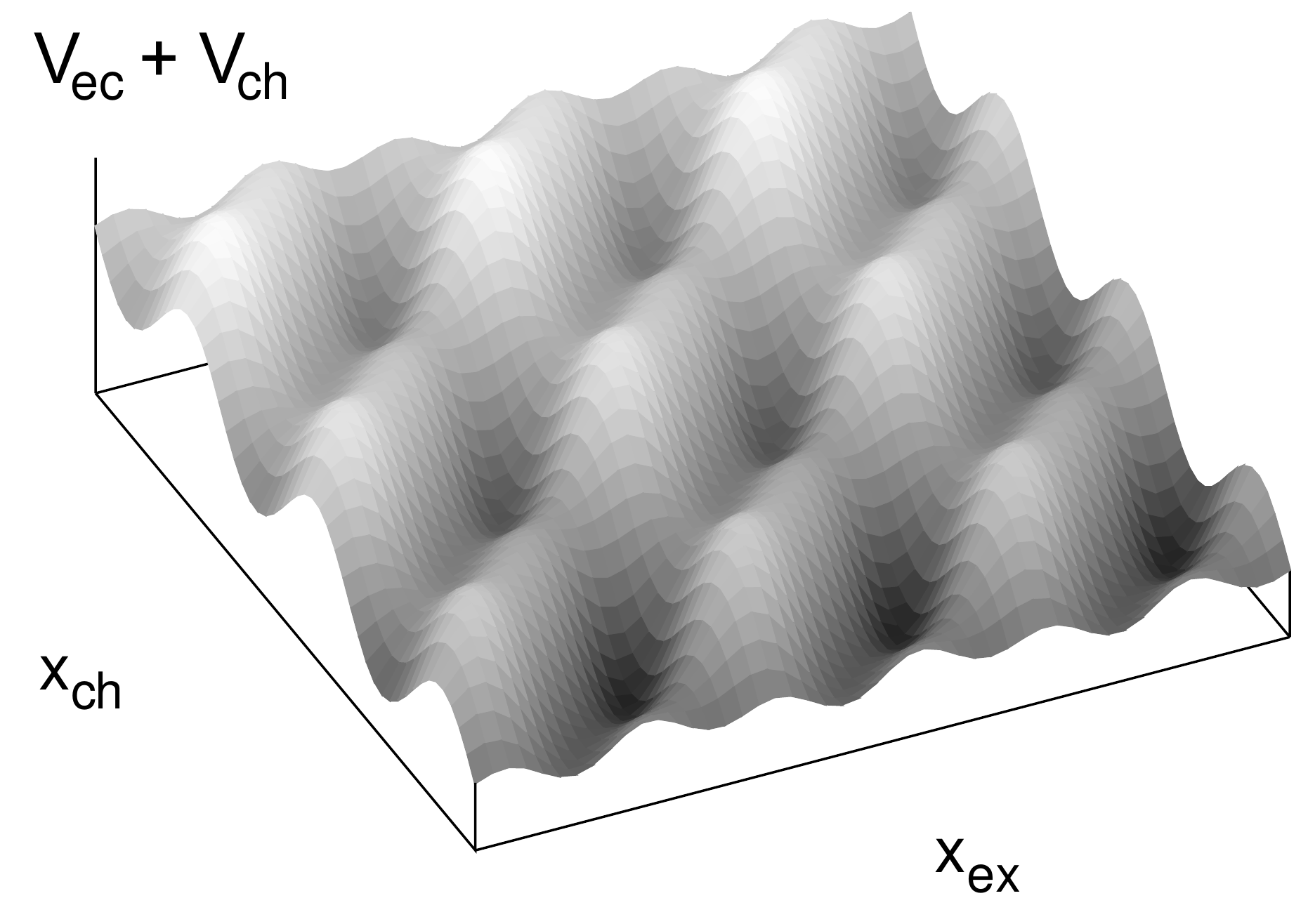}
\caption{Potential (\protect\ref{pottotal}) as a function of
coordinates $x_{\rm ex}$ and $x_{\rm ch}$ for the external and the
chemical freedoms, resp. Greylevel code ranges from black (low) to
white (high).
}\label{potexch}
\end{center}
\end{figure}

To implement these requirements, we compose the full potential as
follows:
\begin{equation} \label{pottotal}
\begin{split}
V(x_{\rm ex},x_{\rm in},x_{\rm ch}) =
V_{\rm ei}(x_{\rm ex},x_{\rm in})+V_{\rm in}(x_{\rm in})+ \\
+ V_{\rm ec}(x_{\rm ex},x_{\rm ch})+V_{\rm ch}(x_{\rm ch}).
\end{split}
\end{equation}
where
\begin{align} \label{potextint}
V_{\rm ei}(x_{\rm ex},x_{\rm in}) =&
-\frac{1+\varepsilon_{\rm ei}\sin(x_{\rm in})}
{1+\varepsilon_{\rm ei}} \notag\\
&\times \frac{\sin(x_{\rm ex})+A\sin(2x_{\rm ex})}{2f_A},\\
V_{\rm in}(x_{\rm in}) =& -K\sin(x_{\rm in}), \label{potint}\\
V_{\rm ec}(x_{\rm ex},x_{\rm ch}) =& -\varepsilon_{\rm ec}\sin(x_{\rm ch})
\sin(x_{\rm ex}-\delta), \label{potextchem}\\
V_{\rm ch}(x_{\rm ch}) =& -\frac{E_{\rm ch}}{2\pi} x_{\rm ch}.
\label{potchem}
\end{align}
The parameter $A$ determines the asymmetry of the ratchet potential
\cite{Fla00}. The coupling between external and internal and between
external and chemical freedoms is controlled, respectively, by
$\varepsilon_{\rm ei}$ and $\varepsilon_{\rm ec}$. We plot the
potential (\ref{pottotal}) as a function of external
and chemical coordinates in Fig.~\ref{potexch}. Diagonal channels
corresponding to a rigid association of one hydrolysis reaction per
spatial step in $x_{\rm ex}$ can be discerned (cf.\ Ref.~\cite{KB00}).

Moreover, to represent the fast internal as well as ambient degrees of
freedom in a collective manner, we include dissipation terms for
$x_{\rm ex}$ and $x_{\rm ch}$ corresponding to velocity-proportional
(Ohmic) friction. Working on the microscopic level, we complement them
by random forces representing thermal noise, with autocorrelation
functions related to dissipation rates through the
fluctuation-dissipation theorem \cite{Rei65}. All in all, we thus
arrive at a set of coupled stochastic differential equations
\cite{Oks98},
\begin{equation} \label{eqsmotion}
\begin{split}
m_{\rm ex} \ddot{x}_{\rm ex} =&
-m_{\rm ex}\gamma_{\rm ex} \dot{x}_{\rm ex} + 
\frac{1+\varepsilon_{\rm ei}\sin(x_{\rm in})}
{1+\varepsilon_{\rm ei}} \\
&\times \frac{\cos(x_{\rm ex})+2A\cos(2x_{\rm ex})}{2f_A} \\
&+\varepsilon_{\rm ec}\sin(x_{\rm ch})\cos(x_{\rm ex}-\delta)
+\xi(t),\\
m_{\rm in} \ddot{x}_{\rm in} =&
\,K\cos(x_{\rm in}) \\
& +\frac{\varepsilon_{\rm ei} \cos(x_{\rm in})}
{1+\varepsilon_{\rm ei}}\,\,
\frac{\sin(x_{\rm ex})+A\sin(2x_{\rm ex})}{2f_A},\\
m_{\rm ch} \ddot{x}_{\rm ch} =&
-m_{\rm ch}\gamma_{\rm ch} \dot{x}_{\rm ch} +
\frac{\Delta V}{2\pi} \\
& +\varepsilon_{\rm ec}\cos(x_{\rm ch}) \sin(x_{\rm ex}-\delta).
\end{split}
\end{equation}
The random force $\xi(t)$ is defined by $\langle \xi(t)\rangle = 0$ and
$\langle \xi(t)\xi(0)\rangle = 2m_{\rm ex} \gamma_{\rm ex}kT\delta(t)$ at
temperature $T$. The parameters $\gamma_{\rm ex}$ and $\gamma_{\rm ch}$
are the friction coefficients for the external and the chemical
freedoms, respectively.

The numerical results presented in the following have been obtained by
solving Eqs.~(\ref{eqsmotion}) by Conventional Brownian Dynamics, see
Refs.~\cite{Oks98,BH98,BH99}. Typical parameter values used in our
simulations are summarized in Tab.~\ref{paramval}.

\begin{table}[h!]
\begin{center}
\begin{tabular}{lllllll}\hline
{\bf Parameter}&$m_{\rm ex}$&$m_{\rm in}$&$m_{\rm ch}$
&$K$&$\varepsilon_{\rm ec}$&$\varepsilon_{\rm ei}$
\\ \hline
{\bf Value}&$1$&$0.1$&$100$ &$0.1$&$1.0$&$1.0$
\\ \hline
\end{tabular}
\vspace*{0.4cm}
\begin{tabular}{lllllll}\hline
{\bf Parameter}&$A$&$kT$&$\Delta V$&$\delta$ &$\gamma_{\rm ex}$
&$\gamma_{\rm ch}$
\\ \hline
{\bf Value}&$0.25$&$1$&$100$&$4.53$&$0.1$&$1$
\\ \hline
\end{tabular}
\caption{Default values for the parameters of our model used wherever
not indicated otherwise.} \label{paramval}
\end{center}
\end{table}

\section{Dynamics and phase-space structure}
\label{dynamics}
The principal parameters determining the behaviour of our model are
$A$ (asymmetry of the ratchet potential), $\varepsilon_{\rm ei}$
(coupling external to internal freedom), $\varepsilon_{\rm ec}$
(coupling external to chemical freedom), $\gamma_{\rm ex}$,
$\gamma_{\rm ch}$ (friction) and temperature $T$. This quite
high-dimensional parameter space is delimited by the following
asymptotics:

\begin{enumerate}
\item \textit{Symmetric external potential} For $A \to 0$, the
simultaneous presence of symmetry-related counterpropagating
trajectory pairs identically cancels transport.
\item \textit{Internal freedom decoupled} In the case $\varepsilon_{\rm
ei} = 0$, the system reduces to a point-particle ratchet, broadly
documented in the literature in all its facets (overdamped,
underdamped with nontrivial dynamics beyond mere relaxation,
Hamiltonian), see \cite{Rei02}.
\item \textit{Chemical freedom decoupled} As we are dealing with an
autonomous system without external driving, decoupling the chemical
freedom deprives the ratchet of its energy source, so that transport
dies out. Even in the Hamiltonian case, a driving is necessary for
directed transport to occur, since there is then no other means to
break time-reversal invariance (apart from magnetic fields which
however play no r\^ole in a biochemical context).
\item \textit{Hamiltonian dynamics} In the limit $\gamma_{\rm
ex},\gamma_{\rm ch} \to 0$ of vanishing friction, the system becomes a
Hamiltonian ratchet with internal freedom. However, in the absence of
an asymmetric external driving or a magnetic field, the system is then
time-reversal invariant and directed currents are excluded.
\end{enumerate}

In the following, we consider in more detail a crucial aspect of the
dynamics, the structure of the attractor and its connectivity along
the lattice as a function of the ``driving force'', i.e., the coupling
to the chemical freedom, as well as friction and noise strength.

\begin{figure}[h!]
\begin{center}
\includegraphics[width=0.4\textwidth]{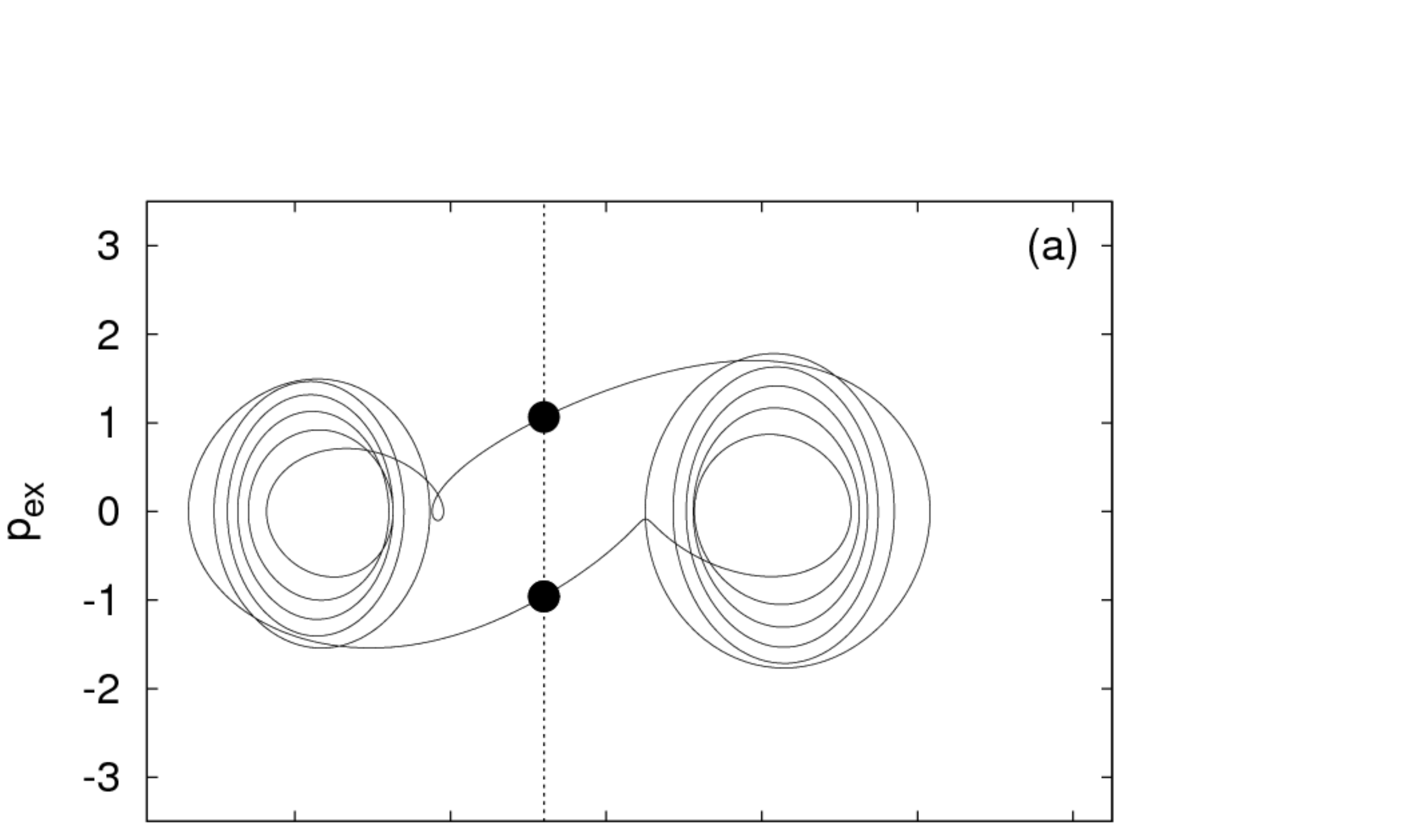}
\includegraphics[width=0.4\textwidth]{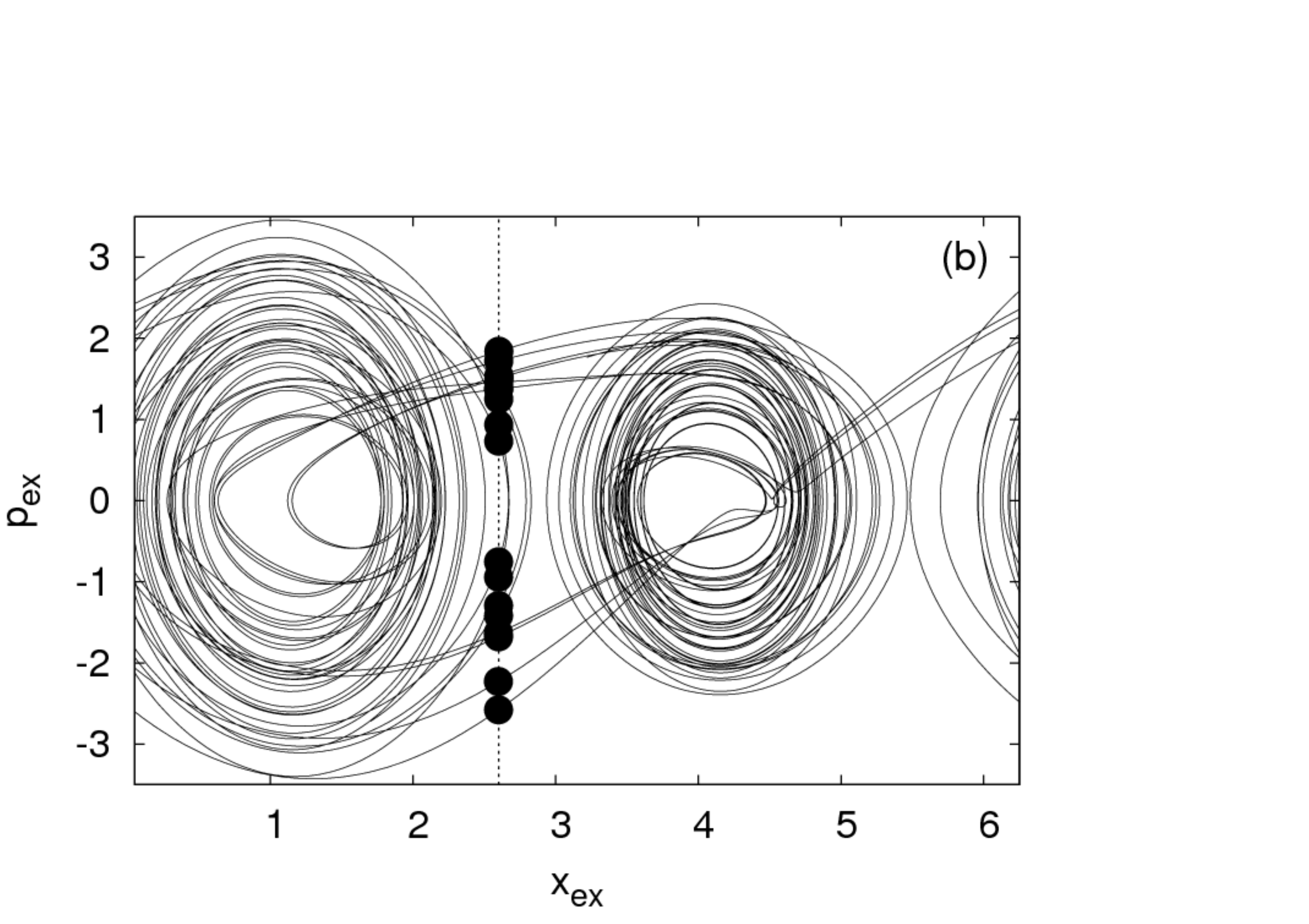}
\caption{Attractor of the system (\protect\ref{eqsmotion}) projected
onto the plane $(p_{\rm ex},x_{\rm ex})$, for $\varepsilon_{\rm ec} =
1.0$ (a) and $7.1$ (b). The dotted line indicates the surface of section
underlying Fig.~\protect\ref{attrachem} below.
}\label{attracphase1}
\end{center}
\end{figure}

\begin{figure}[h!]
\begin{center}
\includegraphics[width=0.4\textwidth]{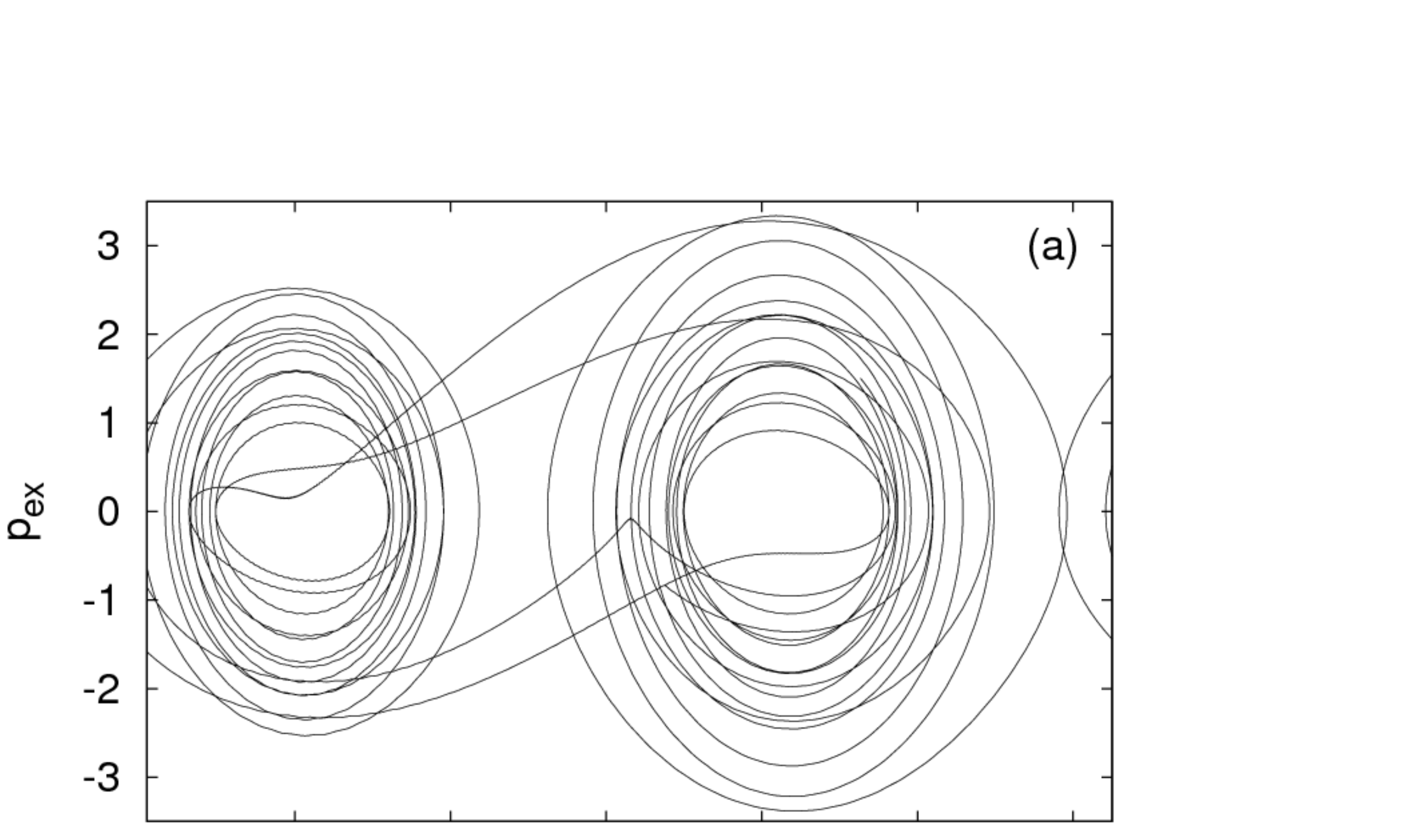}
\includegraphics[width=0.4\textwidth]{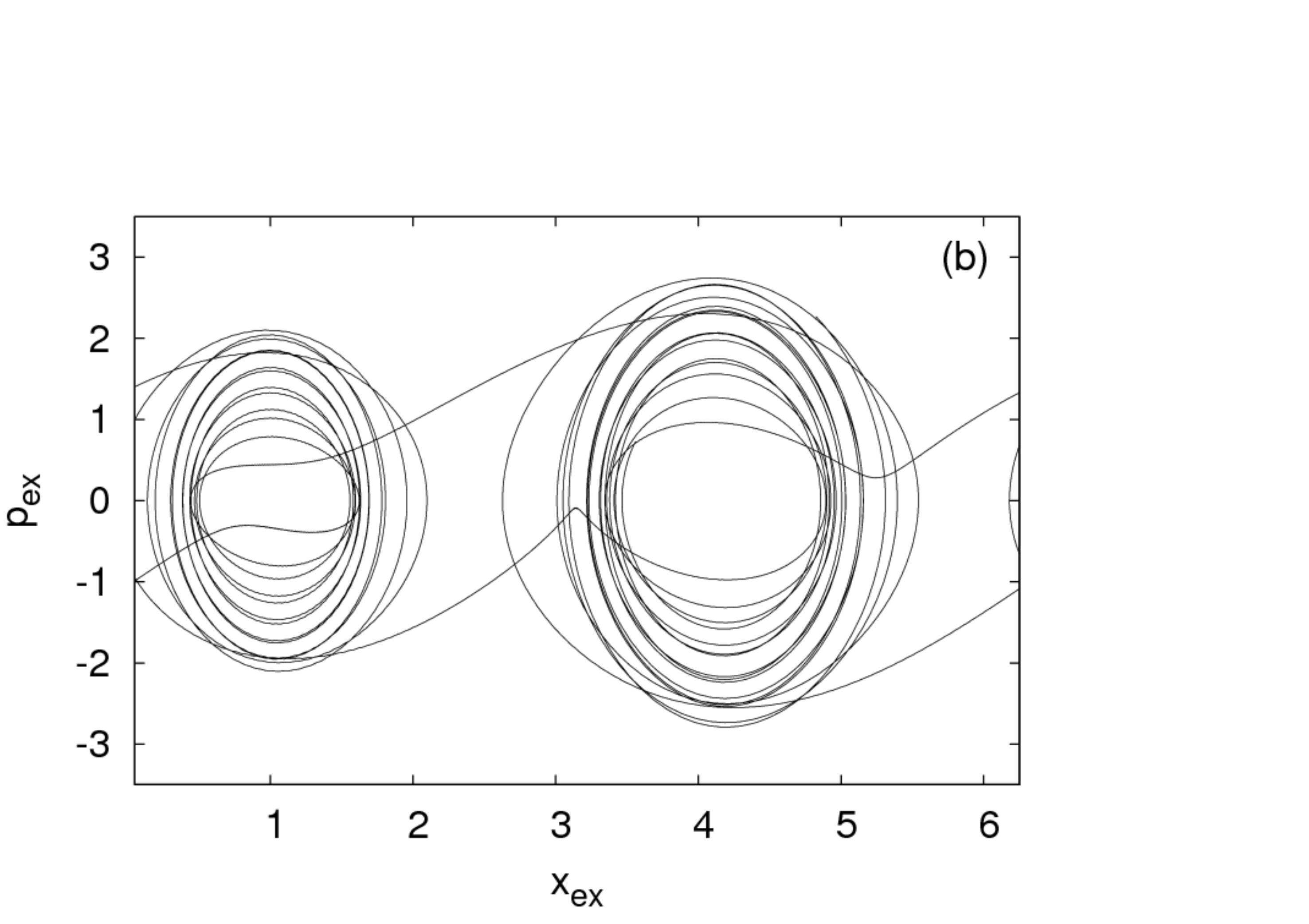}
\caption{As Fig.~\protect\ref{attracphase1} but for $\varepsilon_{\rm
ec} =6.0$ (a) and $2.5$ (b).
}\label{attracphase2}
\end{center}
\end{figure}

Figure \ref{attracphase1} shows the attractor, as a converged
trajectory projected onto the $(p_{\rm ex},x_{\rm ex})$-plane, for
$\varepsilon_{\rm ec} = 1.0$ (panel a) and $\varepsilon_{\rm ec} = 7.1$
(b). While for low coupling we discern a limit cycle, restricted to a
unit cell of the potential, for strong coupling it has given way to a
strange attractor which connects across the unit cell boundaries along
the entire lattice.

These two crucial ``crises'', though, the emergence of a strange
attractor and its change of topology from isolated to conected, are
independent of one another. This is evidenced in Fig.~\ref{attracphase2}a,
analogous to Fig.~\ref{attracphase1} but at intermediate values of the
coupling, where we observe a strange attractor still bounded within a
unit cell. Conversely, Fig.~\ref{attracphase2}b shows a limit cycle
extending periodically along the lattice.

\begin{figure}[h!]
\begin{center}
\includegraphics[width=0.4\textwidth]{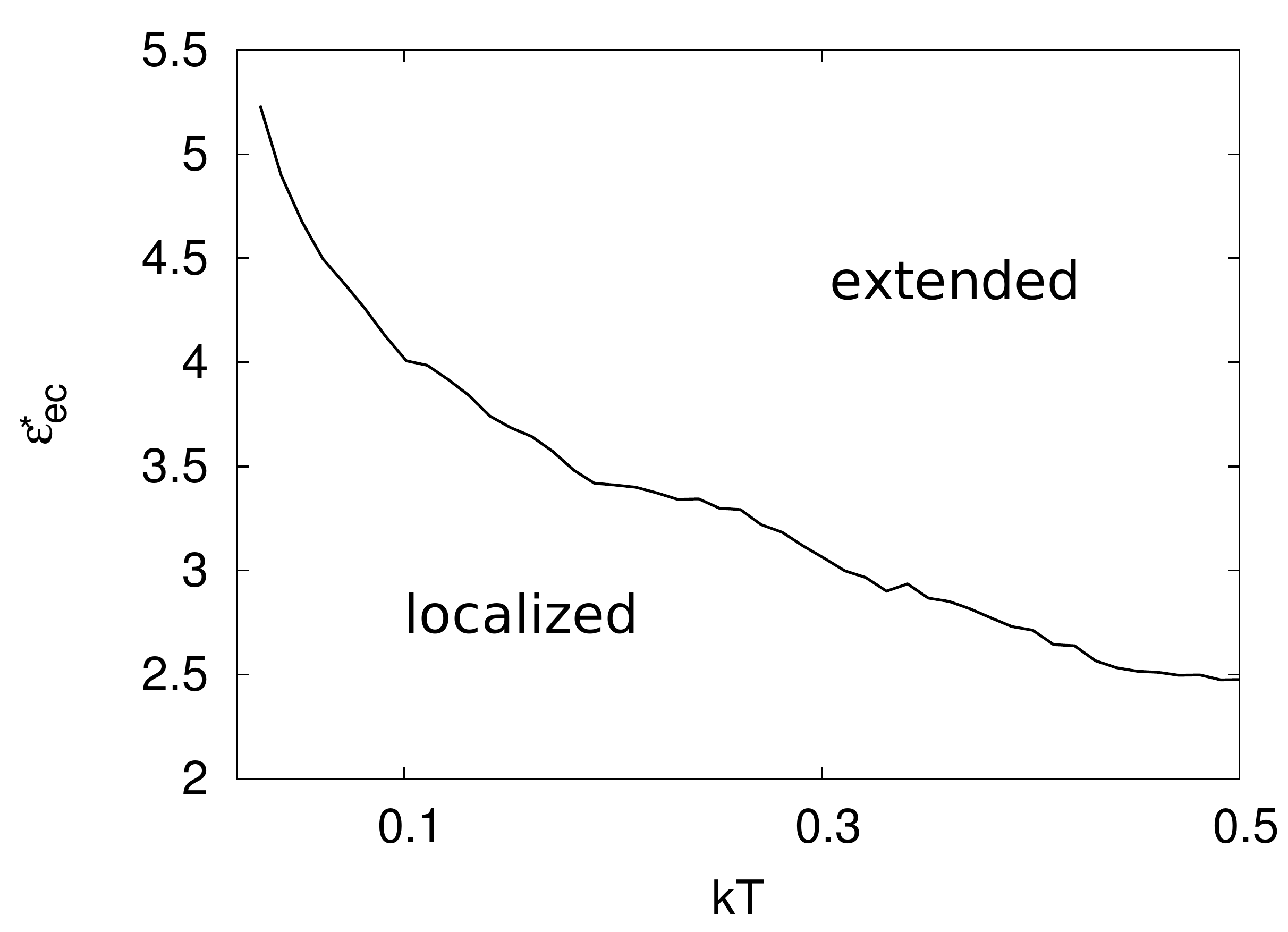}
\caption{Threshold value of $\varepsilon_{\rm ec}^*$ for the
transition from localized to extended attractors as a function of the
bath temperature $T$.
}\label{attracnoise}
\end{center}
\end{figure}

Besides the fundamental r\^ole of thermal fluctuations in directed
transport through the ``rectification of noise'' outside equilibrium,
it affects the dynamics in a more common way by softening attractor
structures and thus contributing to a merger of attractors along the
lattice. Figure \ref{attracnoise} clearly demonstrates how noise tends
to lower the threshold $\varepsilon_{\rm ec}^*$ in terms of the
coupling to the chemical freedom for the transition to transport along
the system.

\section{Directed transport} \label{transport}
\begin{figure}[h!]
\begin{center}
\includegraphics[width=0.4\textwidth, angle=0]{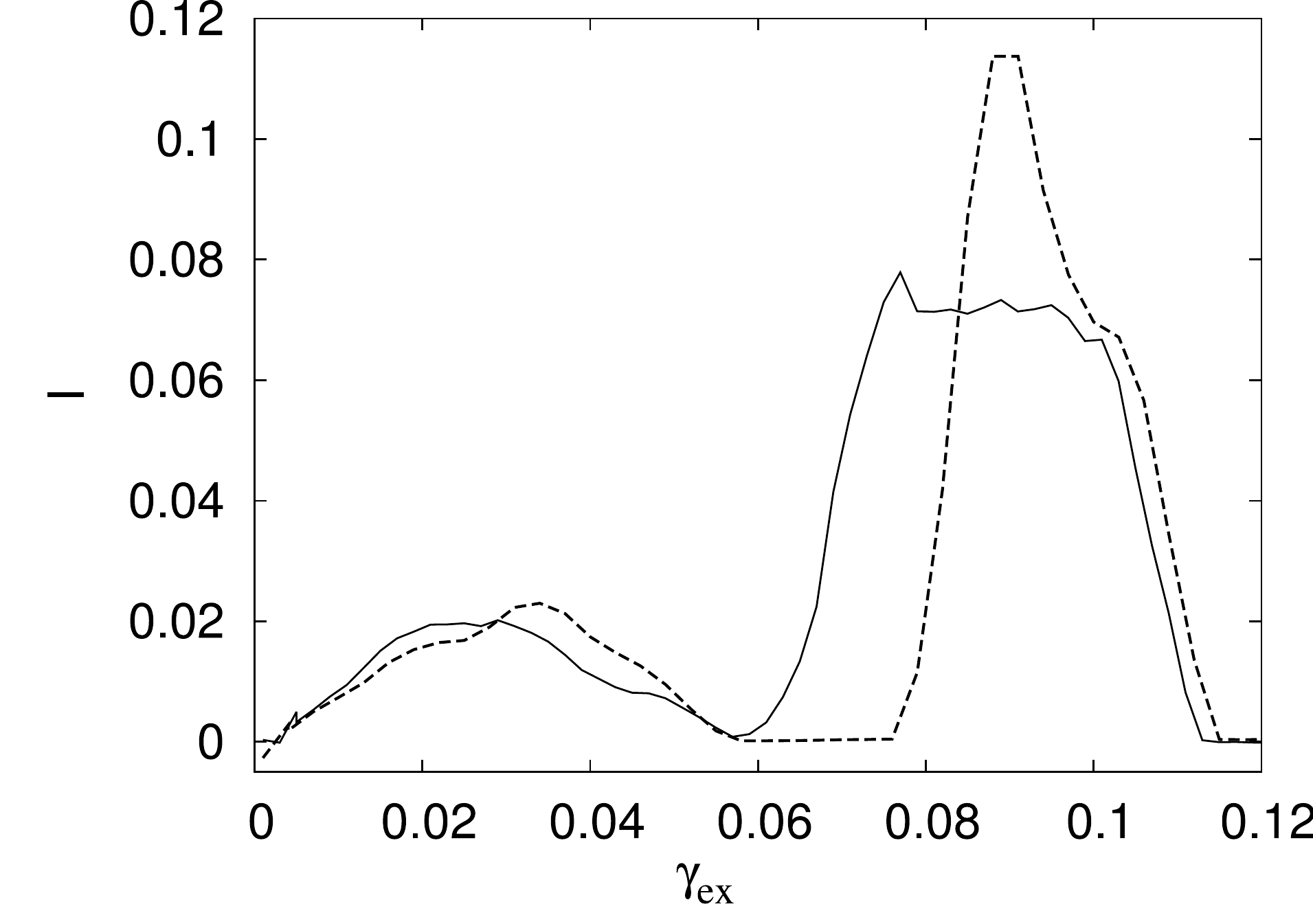}
\caption{Mean current (\protect\ref{current}) as a function of the
friction coefficient $\gamma_{\rm ex}$ for the external degree of
freedom, for $\varepsilon_{\rm ec} = 7.08$ (full line) and 5.35
(dashed).
}\label{currentfric}
\end{center}
\end{figure}

We here measure directed transport in terms of the mean velocity of
the external freedom,
\begin{equation} \label{current}
I=\left<\dot{x}_{\rm ex}\right> =
\lim_{t\to \infty} \frac{1}{t} 
\left<x_{\rm ex}(t)-x_{\rm ex}(0)\right>,
\end{equation}
taking averages, e.g., over initial conditions within the same basin
of attraction or over realizations of a stochastic force.
The generation of non-zero currents in ratchet-like systems
depends on the following independent general necessary conditions:
\renewcommand{\theenumi}{\roman{enumi}}
\begin{enumerate}
\item \label{asymmetry} The breaking of all binary symmetries
involving an inversion of momentum, such as in particular time
reversal and parity \cite{Fla00}.
\item \label{rotation} The existence of trajectories extending over
different unit cells of the potential, or, if the external freedom
is cyclic, corresponding to rotational as opposed to librational
motion.
\end{enumerate}
Concerning condition (\ref{asymmetry}), in the absence of an external driving
or a coupling to a magnetic field, it is only the friction terms in
the equations of motion which break time-reversal invariance. It is
evident from Fig.~\ref{currentfric} that there is no transport for
vanishing friction $\gamma_{\rm ex} = 0$, and likewise for vanishing
asymmetry parameter $A = 0$ (not shown).

As to (\ref{rotation}), in the presence of friction, the existence of
persistent transporting trajectories is not guaranteed, in difference
to the case of Hamiltonian ratchets where there is always an asymptotic
regime of quasi-free motion at high energies \cite{SDK05}. As discussed
above, the topology of attractors depends sensitively on couplings,
friction parameters, noise strengths etc. In the subsequent paragraphs,
we consider in detail the influence of a these parameters on the
generation of directed currents in the system.

\subsection{Resonance between chemical and external
freedoms} \label{resonance}
\begin{figure}[h!]
\begin{center}
\includegraphics[width=0.4\textwidth, angle=0]{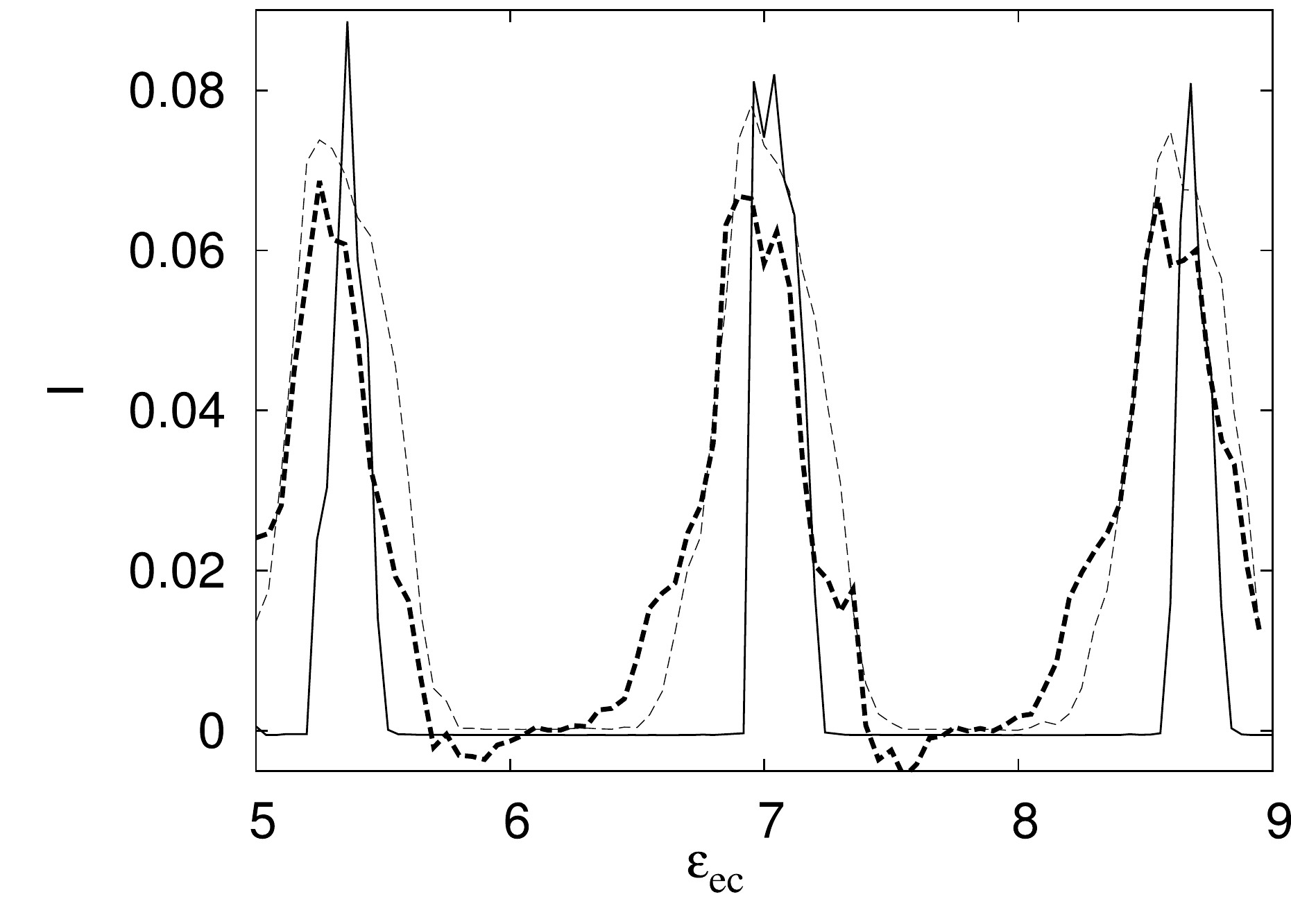}
\caption{Mean current (\protect\ref{current}) as a function of the
coupling $\varepsilon_{\rm ec}$ between external and chemical freedom,
for different values of the coupling between external and internal
freedom, $\varepsilon_{\rm ei} = 0$ (full line), $0.5$ (bold dashed),
$1$ (light dashed).
}\label{currentchem}
\end{center}
\end{figure}

\begin{figure}[h!]
\begin{center}
\includegraphics[width=0.4\textwidth, angle=0]{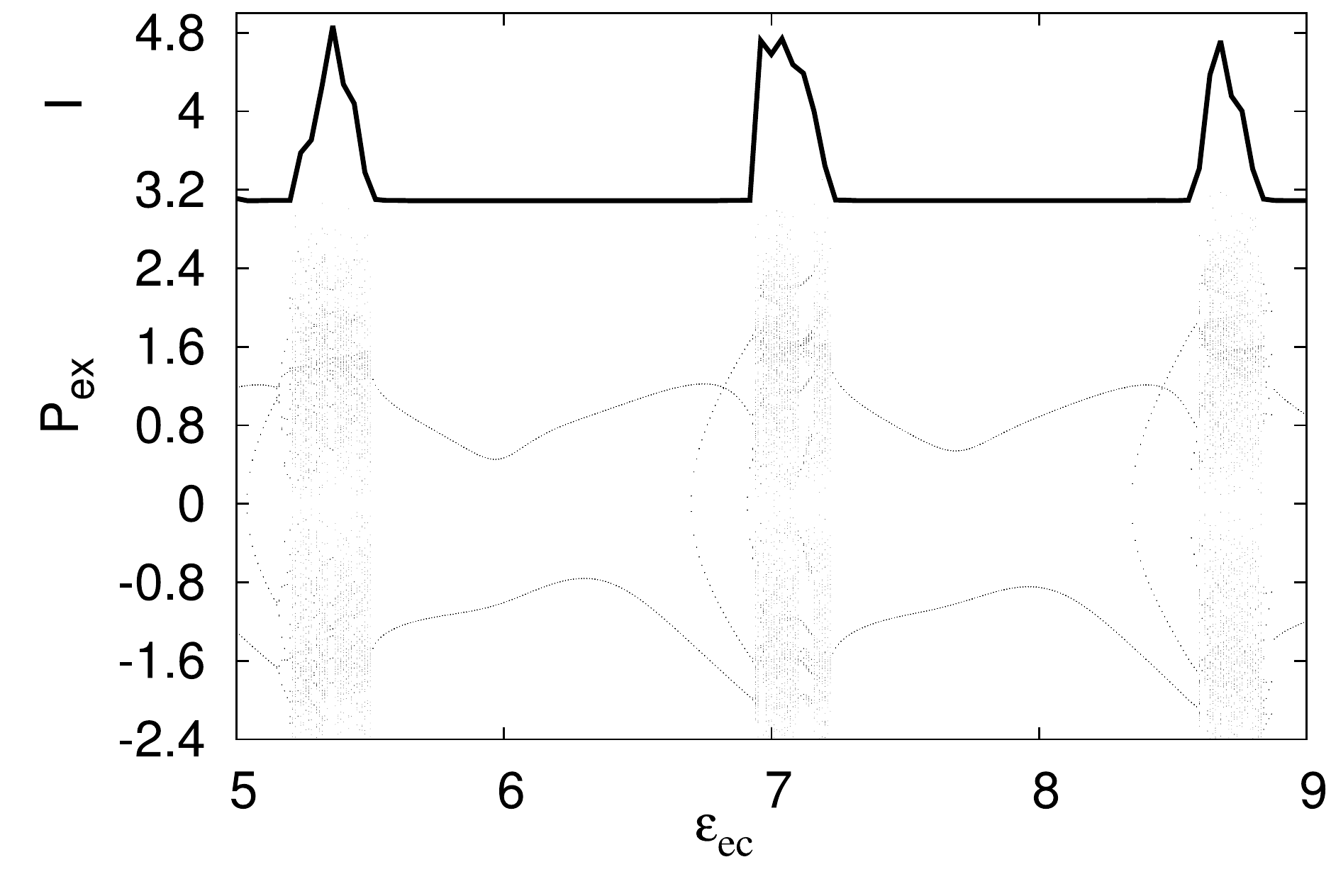}
\caption{Attractor of the system (\protect\ref{eqsmotion}) in terms of
the surfaces of section indicated in Figs.~\protect\ref{attracphase1}
and \protect\ref{attracphase2} as a function of the
coupling $\varepsilon_{\rm ec}$. Upper part: Mean current
(\protect\ref{current}) (bold dashed) as a function of the same
parameter.
}\label{attrachem}
\end{center}
\end{figure}

In the absence of an external periodic driving, our model is lacking a
precise clock that would lead to clearcut resonances. The
high inertia of the chemical freedom, however, and the
corresponding relatively regular injection of energy ``quanta'' also
defines an approximate time scale that gives rise to resonance-like
phenomena. This is obvious from Fig.~\ref{currentchem}, where marked
peaks of the current occur at approximately equidistant values of the
external-to-chemical coupling, $\varepsilon_{{\rm ec},n} = n
\varepsilon_{{\rm ec},1}$, $n = 1,2,\ldots$, with $\varepsilon_{{\rm
ec},1} \approx 1.8$ in this case. For this value, there is a 1:1
resonance between the two coupled freedoms, i.e., in terms of
Fig.~\ref{potexch}, the system is moving along diagonal valleys in the
two-dimensional array of potential maxima.

However, this resonant behaviour does not amount to simple
oscillatory motion in $x_{\rm ex}$. As is evident from
Fig.~\ref{attrachem}, where we plot a section of the attractor,
cf.\ Fig.~\ref{attracphase1}, vs.\ $\varepsilon_{\rm ec}$, the peaks of
the current coincide precisely with intervals in $\varepsilon_{\rm
ec}$ where the attractor is fractal. What is more, the sharp
definition without shoulders of these peaks can be related to abrupt
changes in the attractor structure from limit cycle to chaotic and
back.

\subsection{Effects of noise} \label{noise}
\begin{figure}[h!]
\begin{center}
\includegraphics[width=0.4\textwidth, angle=0]{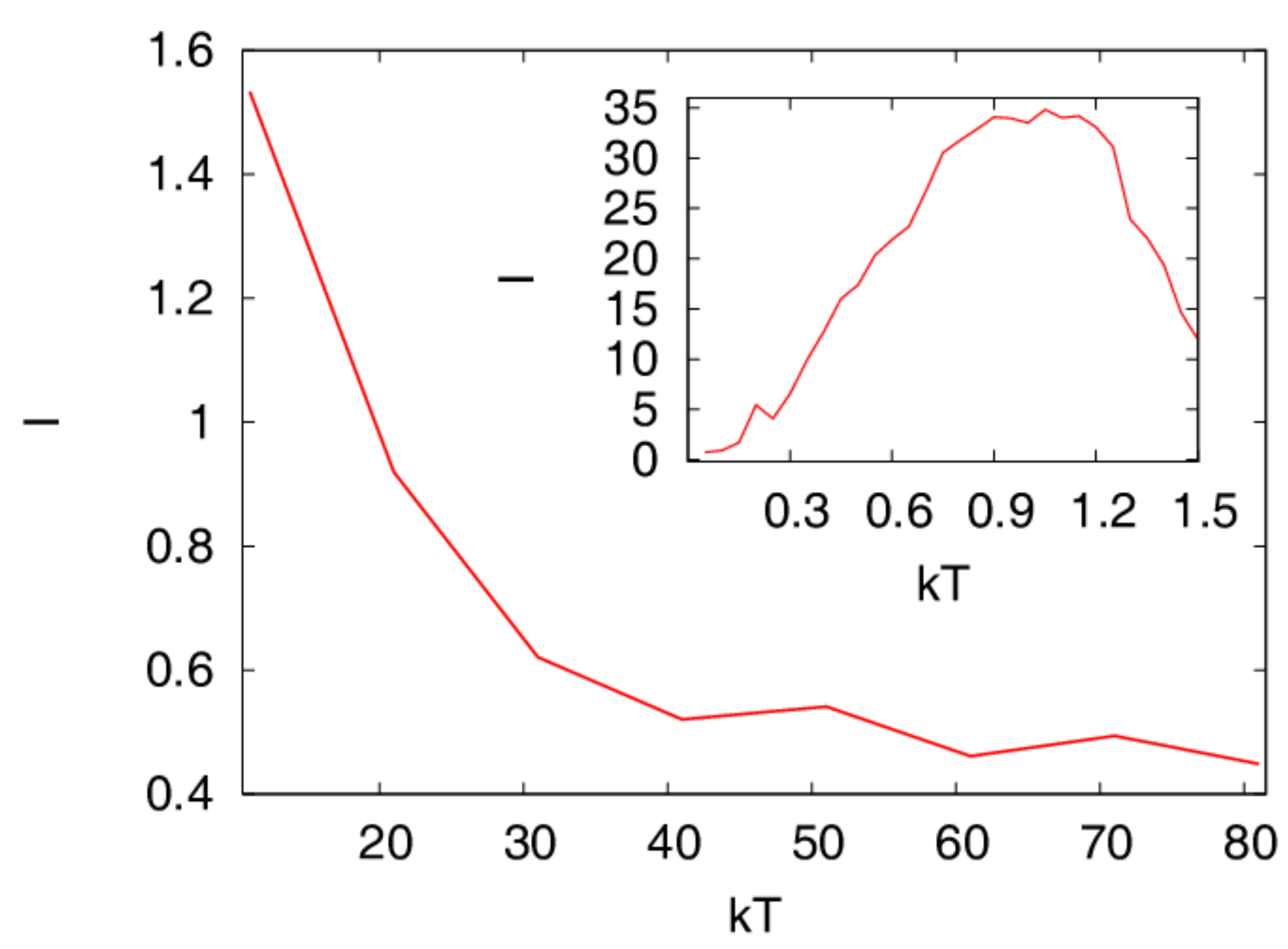}
\caption{Mean current (\protect\ref{current}) as a function of the
bath temperature $T$. Inset: enlargement of the low-temperature regime
$0 \leq T \leq 1.5$.
}\label{currenttemp}
\end{center}
\end{figure}

Figure \ref{currenttemp} shows the current vs.\ noise strength in
terms of bath temperature $T$. As would be expected, too strong noise
washes out the structure of the attractor, in particular its
asymmetry, a necessary condition for directed transport to occur. The
inset reveals a more interesting phenomenon: For the other parameter
values in question kept constant, there is no extended attractor in
the absence of thermal fluctuations, so that the current vanishes at
$T = 0$. As a consequence, there exists a maximum, albeit broad, of
the current at a bath temperature $T \approx 1$, in analogy to
stochastic resonance \cite{GH&98}.

\subsection{R\^ole of the internal freedom} \label{internal}
\begin{figure}[h!]
\begin{center}
\includegraphics[width=0.4\textwidth]{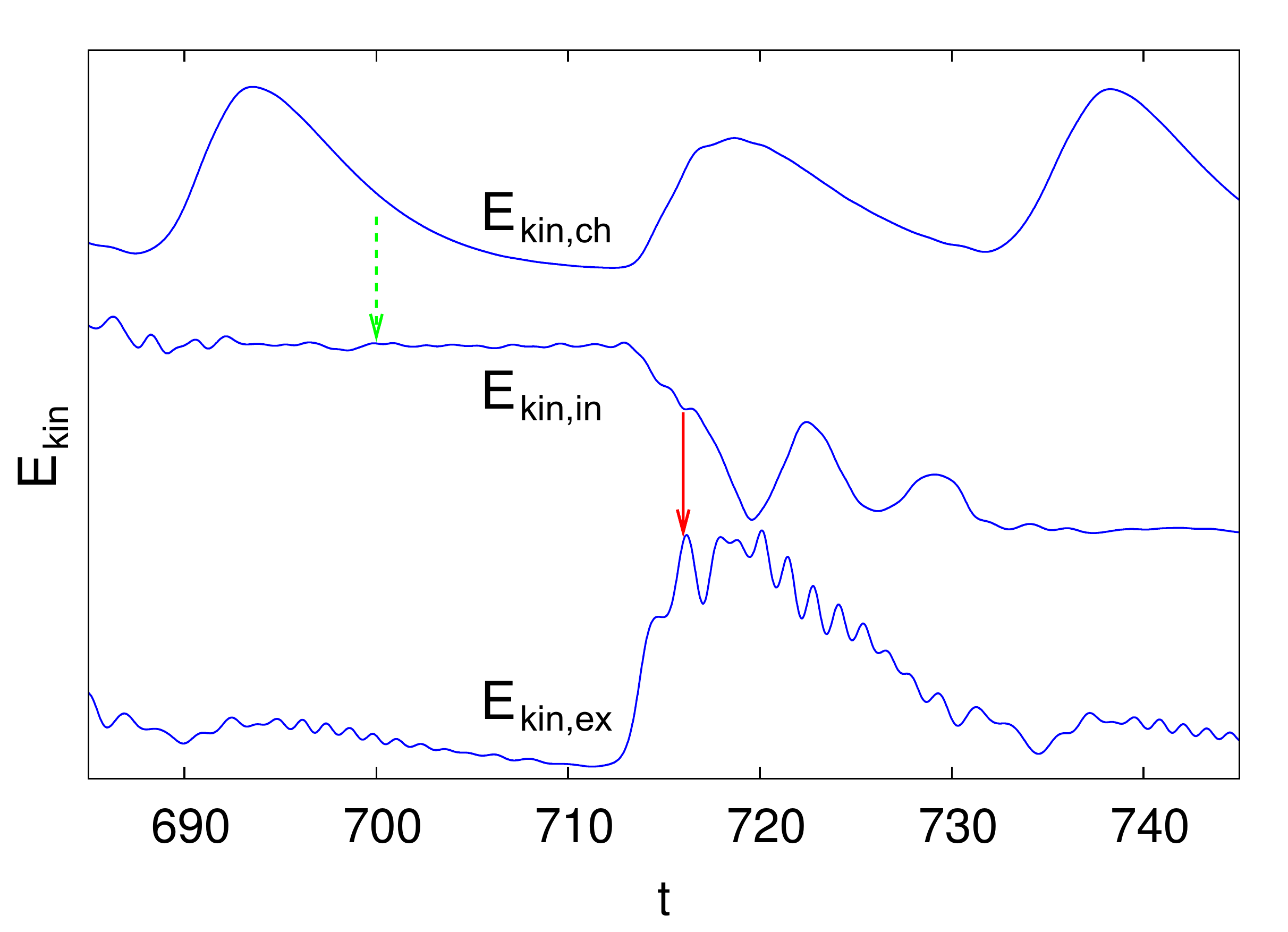}
\caption{Kinetic energy contained in the chemical (upper graph),
internal (middle) and external (lower graph) freedoms as a function of
time. The figure shows a particular case. Time dependence is slightly
smoothed to suppress irrelevant fluctuations. Arrows indicate typical
energy-transfer events from chemical to internal (dashed, green) and
from internal to external freedom (full, red). Small-scale
oscillations in the lower graph correspond to unit transport steps in
$x_{\rm ex}$. See text.
}\label{entrans}
\end{center}
\end{figure}

A first indication of the influence of the internal freedom can be
observed in Fig.~\ref{currentchem}, comparing the width of the current
peaks for different values of the external-internal coupling. They
systematically grow broader with increasing
$\varepsilon_{\rm ei}$. This means that off resonance, directed
transport is supported or even enabled in the first place by the
presence of the internal freedom.

Further insight into the underlying mechanism is provided by
Fig.~\ref{entrans}, where we plot the kinetic energy in the three
respective freedoms as a function of time. It shows a clear tendency
for a temporary storage of energy in the internal freedom: The system
receives energy  almost periodically and in nearly equal portions from
the chemical freedom. It tends to accumulate in the internal freedom
(green arrow) till a threshold is reached, whereupon it is discharged
(red arrow) almost completely to the external freedom, giving rise to
a burst of transport steps. This ``toilet-flush'' behaviour is
well-known from neuron firing \cite{HH52}.

Comparing with Fig.~\ref{attrachem}, we observe moreover that the
onset of current with increasing  $\varepsilon_{\rm ei}$ also
coincides with a structural change of the attractor from limit cycle
to strange. To be sure, two autonomous degrees of freedom are
sufficient to allow for chaotic behaviour. Notwithstanding, we can
conclude that another relevant r\^ole of the internal freedom is to 
induce a richer, more stochastic dynamics of the system.

\section{Conclusion} \label{end}
The present work pretends to reduce the gap between abstract models of
directed transport and detailed studies of specific motor molecules by
constructing a system that endows ratchets with a few essentials of
combustion motors, an internal degree of freedom and a coupling to an
autonomous chemical freedom as energy source instead of a
periodic external force, but ignoring the specifics of particular
(classes of) molecules. This allows us to analyze in
general terms the r\^ole of the internal freedom in the generation of
transport. We find convincing evidence that it serves as an interface
which effectively decouples the output of mechanical work in the form
of transport steps from the energy input through hydrolysis or similar
reactions. We can observe directly how energy is being stored
temporarily in the internal degree of freedom before being channeled
to the external freedom. In addition, the internal freedom induces 
transitions towards a more irregular dynamics of the external freedom, 
which also contributes to the generation of directed currents.

These results direct attention to processes of energy transfer and
dissipation within the molecule, and in particular indicate the
functional advantage of a cascaded energy redistribution through
intermediate steps \cite{Gar97}, refining the simple picture of an
external  (transport) degree of freedom coupled directly to a
structureless bath. Even within the scope of our three-dimensional
model, such questions could be studied by changing the configuration
of couplings, e.g., coupling the external to the chemical freedom only
through the internal one instead of internal--external--chemical as in
the present work.

Subtler effects of noisy nonlinear dynamics like stochastic resonance 
\cite{GH&98} have not been considered in depth. In the context of our
model, it is expected to occur in a parameter regime of weak noise
close to a resonance of the chemical with the external freedom, cf.\ 
Sec.~\ref{resonance}, lowering the threshold to directed transport as 
a function of the coupling to the chemical freedom. Quantum effects 
have not been taken into account, either. We expect relevant 
modifications of the picture, if any, for the dynamics of the chemical 
degree of freedom and the injection of chemical energy into functional 
modes where the discretization of energy and quantum coherence might 
play a major r\^ole.

\section*{Acknowledgements}
We would like to thank Camilo Aponte, Alfonso Leyva, and Jos\'e Daniel 
Mu{\~n}oz for inspiring discussions and valuable bibliographical 
information. One of us (TD) gratefully acknowledges financial support 
by Colciencias and by Volkswagen Foundation  during preparation of this 
work and thanks for the hospitality extended to him by Rensselaer 
Polytechnic Institute (Troy, NY, USA).








\end{document}